\documentclass[runningheads,a4paper]{llncs}
\usepackage{makeidx}  
\usepackage{graphicx}
\begin{document}
\mainmatter              
\title{An Integrated Research Infrastructure for Validating Cyber-Physical Energy Systems}
\titlerunning{An Integrated RI for Validating CPES}  
\author{
	T.~I.~Strasser\inst{1} \and
	C.~Moyo\inst{1} \and
	R.~Br\"undlinger\inst{1} \and	
	S.~Lehnhoff\inst{2} \and
	M.~Blank\inst{2} \and	
	P.~Palensky\inst{3} \and
	A.~A.~van~der~Meer\inst{3} \and
	K.~Heussen\inst{4} \and
	O.~Gehrke\inst{4}
	J.~E.~Rodriguez\inst{5} \and
	J.~Merino\inst{5} \and
	C.~Sandroni\inst{6} \and
	M.~Verga\inst{6} \and
	M.~Calin\inst{7} \and
	A.~Khavari\inst{7} \and
	M.~Sosnina\inst{7} \and 
    E.~de~Jong\inst{8} \and
    S.~Rohjans\inst{9} \and 
    A.~Kulmala\inst{10} \and 
    K.~M\"aki\inst{10} \and 
    R.~Brandl\inst{11} \and 
    F.~Coffele\inst{12} \and 
    G.~M.~Burt\inst{12} \and
    P.~Kotsampopoulos\inst{13} \and
    N.~Hatziargyriou\inst{13}
}
\authorrunning{T.~I.~Strasser et al.} 
\institute{
	AIT Austrian Institute of Technology, Vienna, Austria \and
	OFFIS e.V., Oldenburg, Germany \and
	Delft University of Technology, Delft, The Netherlands \and
	Technical University of Denmark, Lyngby, Denmark \and
	TECNALIA Research \& Innovation, Derio, Spain \and
	Ricerca sul Sistema Energetico, Milano, Italy \and
	European Distributed Energy Resources Lab. (DERlab) e.V., Kassel, Germany \and 
    DNV GL, Arnhem, The Netherlands \and 
    HAW Hamburg University of Applied Sciences, Hamburg, Germany \and
    VTT Technical Research Centre of Finland, Espoo, Finland \and 
    Fraunhofer Inst. of Wind Energy and Energy System Technology, Kassel, Germany \and 
    University of Strathclyde, Glasgow, United Kingdom \and 
    National Technical University of Athens, Athens, Greece \\ [0.15cm]
	\email{thomas.strasser@ait.ac.at}
}
\maketitle              
%
%
\begin{abstract}
Renewables are key enablers in the plight to reduce greenhouse gas emissions and cope with anthropogenic global warming. The intermittent nature and limited storage capabilities of renewables culminate in new challenges that power system operators have to deal with in order to regulate power quality and ensure security of supply. At the same time, the increased availability of advanced automation and communication technologies provides new opportunities for the derivation of intelligent solutions to tackle the challenges. Previous work has shown various new methods of operating highly interconnected power grids, and their corresponding components, in a more effective way. As a consequence of these developments, the traditional power system is being transformed into a cyber-physical energy system, a smart grid. Previous and ongoing research have tended to mainly focus on how specific aspects of smart grids can be validated, but until there exists no integrated approach for the analysis and evaluation of complex cyber-physical systems configurations. This paper introduces integrated research infrastructure that provides methods and tools for validating smart grid systems in a holistic, cyber-physical manner. The corresponding concepts are currently being developed further in the European project ERIGrid.
\end{abstract}
%
%
\section{Introduction}
\label{sec:introduction}

Future power systems have to integrate a higher amount of distributed, renewable energy resources in order to cope with a growing electricity demand, while at the same time trying to reduce the emission of greenhouse gases \cite{IPCC:2014}. In addition, power system operators are nowadays confronted with further challenges due to the highly dynamic and stochastic behaviour of renewable generators (solar, wind, small hydro, etc.) and the need to integrate controllable loads (electric vehicles, smart buildings, energy storage systems, etc.). Furthermore, due to ongoing changes to framework conditions and regulatory rules, technology developments (development of new grid components and services) and the liberalization of energy markets, the resulting design and operation of the future electric energy system has to be altered. 

Sophisticated (systems and component) design approaches, intelligent information and communication architectures, and distributed automation concepts provide ways to cope with the above mentioned challenges and to turn the existing power system into an intelligent entity, that is, a ``Cyber-Physical Energy System (CPES)'' (also known as ``Smart Grid'') \cite{IEA:2011,Farhangi:2010,Strasser:2013}. 

While reaping the benefits that come along with intelligent solutions, it is, however, expected that due to the considerably higher complexity of such solutions, validation and testing will play a significantly larger role in the development of future technology. As it stands, the first demonstration projects for smart grid technologies have been successfully completed, it follows that there is a high probability of key findings and achieved results being integrated in new and existing products, solutions and services of manufacturers and system integrators. Up until now, the proper validation and testing methods and suitably corresponding integrated Research Infrastructure (RI) for smart grids is neither fully available nor easily accessible \cite{Strasser:2017}.  

The aim of this paper is to introduce an approach for integrated RI with corresponding CPES-based system-level validation methods that are being currently implemented in the framework of the European project ERIGrid \cite{ERIGrid:URL}. 

The remaining parts of the paper are organized as follows: Section~\ref{sec:cyber-physical_energy_systems} provides a brief overview of CPES challenges whereas in Section~\ref{sec:future_research_needs} the corresponding research needs are outlined. The concept of the ERIGrid integrated RI is introduced in the following Section~\ref{sec:integrated_research_infrastructure}. The paper concludes with a discussion about and an outlook on future developments.
%
%
\section{Higher Complexity in Cyber-Physical Energy Systems}
\label{sec:cyber-physical_energy_systems}

Smart grid systems usually lead to an increased level of complexity within system operation and management as briefly outlined in the introduction. There is an urgent need for the system flexibility to also be increased, in order to avoid dramatic consequences. It also follows that advanced Information and Communication Technology (ICT), distributed automation approaches and power electronic-based grid components are necessary in order to allow a number of important system functionalities (e.g., power/energy management, demand side management, ancillary services) \cite{IEA:2011,Farhangi:2010}. As a consequence of these developments (distributed) intelligence is needed on four different levels in smart grid systems as outlined in Table~\ref{tab:cpes-levels} \cite{Strasser:2015,Strasser:2017}.

\begin{table}[!htbp]
	\centering
	\caption{CPES -- intelligence on different levels}
	\label{tab:cpes-levels}
	\setlength{\tabcolsep}{0.5em} 
	{\renewcommand{\arraystretch}{1.25}
	\begin{tabular}{c|p{9.50cm}}
		\hline
		\textbf{Level} & \multicolumn{1}{c}{\textbf{Implemented Intelligence}} \\
		\hline
		\hline 
		\textit{System} & System-wide approaches like power utility automation, coordinated voltage control, demand-side management, energy management, etc. are usually executed in a coordinated way on this level, but also factoring in services of the underlying sub-systems and components. Central or distributed control approaches can be are applied. \\
		\hline 
		\textit{Sub-System} & On this layer the control of the underlying sub-systems or components is carried out. Usually the corresponding functions, services, and algorithms have to deal with a limited amount of components (renewable sources, energy storage system, electric vehicle supply equipment, etc.). Micro-grid control or building energy management are representative examples for this layer. Distributed automation architectures are commonly used. \\ 
		\hline 
		\textit{Component} & Nowadays, new components like Distributed Energy Resources (DER), energy storages, electric vehicle supply equipment, or tap-changing transformers providing ancillary services are installed on this layer. Intelligence on this level is either used for local optimization purposes (component behaviour) or for the optimisation of systems/sub-systems on higher levels in a coordinated manner. \\ 
		\hline 
		\textit{Sub-Component} & On this level intelligence is mainly used to improve  local component behaviour (harmonics, flicker, etc.). Power electronics and the necessary corresponding advanced control algorithms are the main drivers for local intelligence. Component controllers can be considered as examples for sub-components. \\ 
		\hline 
	\end{tabular}} 
\end{table}

The top requirements for the realisation of advanced ICT/automation systems and component controllers include flexibility, adaptability, scalability, and autonomy. Open interfaces that support interoperability are also necessary in enabling the above described behaviour \cite{Strasser:2015,Strasser:2013,Strasser:2017}. As a consequence, the electric power system is moving towards a complex cyber-physical system of systems. Not only the design and implementation, but also the validation and deployment of these systems is associated with increasing both the engineering complexity and total life-cycle costs.

In order to address challenges in CPES, such as network limitations, CPES modeling and computational prediction of system uncertainty \cite{macana2011}, multidisciplinary teams that understand the different aspects of CPES and from all layers are needed. CPES research requires control system engineers, engineers familiar with the physical process being controlled (electric generation, electric distribution), communication engineers, and security engineers \cite{morris2009}. The team needs to be assembled based on the particular use case (e.g., modeling/simulation \cite{al2014,ilic2008m,palensky2014}, security \cite{khan2013,sridhar2012}, smart houses/buildings \cite{kleissl2010,wu2011}).  
%
%
\section{Open Issues and Future Research Needs}
\label{sec:future_research_needs}

\begin{figure}[!htbp]
	\centering
	\includegraphics[width=0.82\columnwidth]{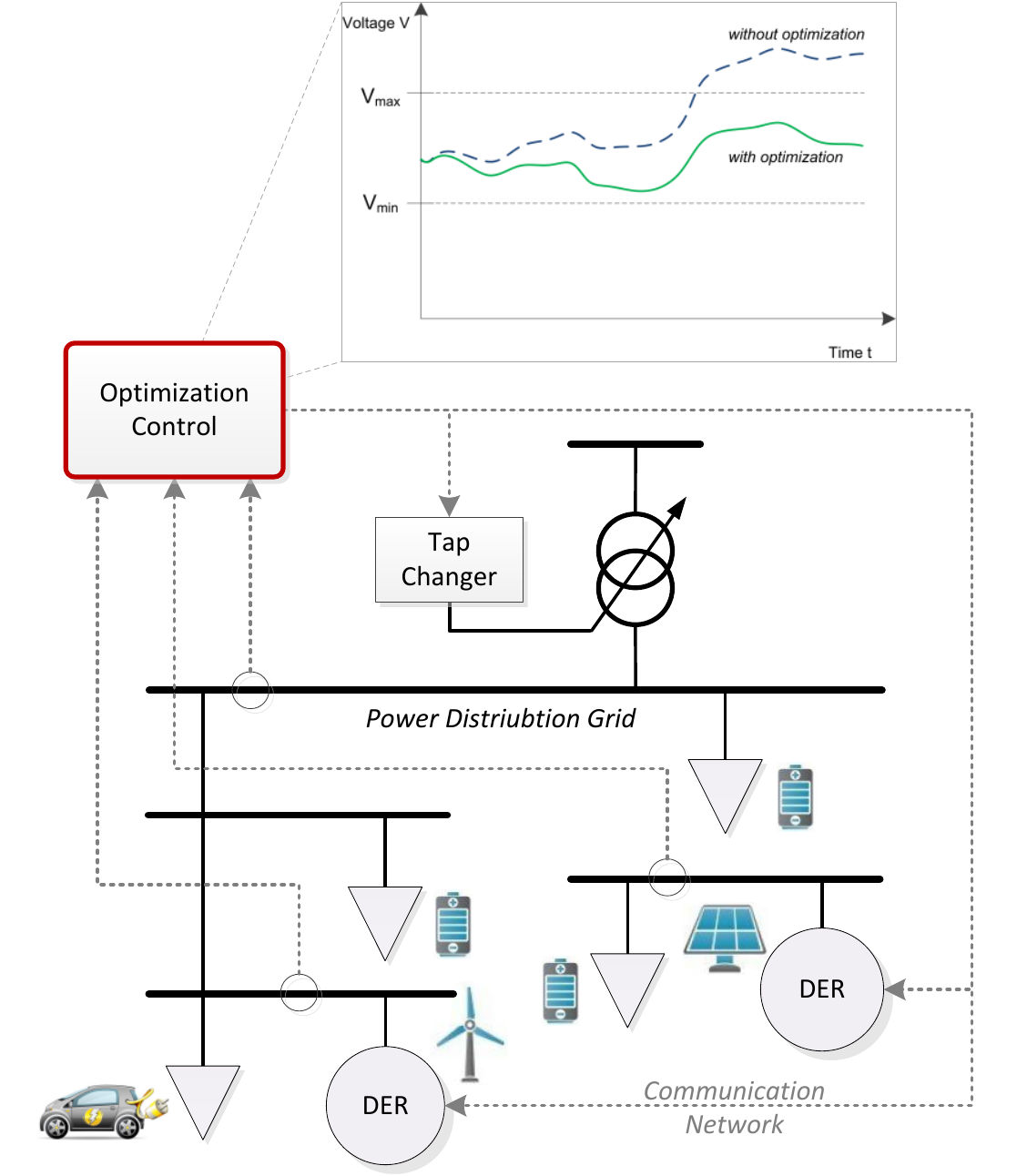}
	\caption{CPES example -- voltage control in an active power distribution grid}
	\label{fig:example}
\end{figure}

To facilitate in the understanding of future CPES validation needs, an illustrative example will be introduced. Figure~\ref{fig:example} shows a coordinated voltage control application in an active power distribution grid. Reactive and active power control provided by DER and electric storage units, together with an On-Load Tap Changing (OLTC) transformer, are used to keep the voltage in the grid in defined boundaries \cite{Stifter:2011}. The control application has to calculate the optimal position of the OLTC and to derive set-points for reactive and active power which is communicated over a communication network to the DER and storage devices.  

In order to guarantee the safe and secure operation of this CPES application various tests need to be carried out before installing it in the field. This includes the validation of the different components (including local control approaches and communication interfaces for the DER, storage, and OLTC devices) on the sub-component and component levels. Nevertheless, the integration of all components and sub-systems is also still one of the most important issues. The proper functionality of all components is not a guarantee that the whole system will behave as expected. A system-level validation of the actual behaviour is necessary in order to prove that the whole CPES application, together with the ICT devices, works properly. 

Up to now, there is no integrated approach for analysing and evaluating smart grid configurations addressing power system, as well as information, communication and automation/control topics that is available \cite{Strasser:2017}. The integration of cyber-security and privacy issues is also not sufficiently addressed by existing solutions. In order to guarantee a sustainable and secure supply of electricity in a smart grid system, with considerably higher complexity and also support the expected forthcoming large-scale roll out of new technologies, a proper integrated RI for smart grid systems is necessary \cite{Strasser:2017}. Such an infrastructure has to support system analysis, evaluation and testing issues. Furthermore, it would foster future innovations and technical developments in the field. 

In summary, the following open issues have been identified and need to be addressed in future research and development \cite{EUR:2011,IEA:2013,SG:2012,Brunner:2016,Strasser:2017}:

\begin{itemize}
	\item A cyber-physical, multi-domain approach for analysing and validating CPES on the system level is missing today; existing methods are mainly focusing on the component level –- system integration topics including analysis and evaluation are not yet addressed in a holistic manner. 
	\item A holistic validation framework (incl. analysis and evaluation/benchmark criteria) and the corresponding RI with proper methods and tools needs to be developed. 
	\item Harmonized and standardized evaluation procedures need to be developed.
	\item Well-educated professionals, engineers and researchers that understand smart grid systems in a cyber-physical manner need to be trained on a broad scale.
\end{itemize}
%
%
\section{ERIGrid Smart Grid Research Infrastructure}
\label{sec:integrated_research_infrastructure}
    
In order to tackle the above aforementioned research needs, a Pan-European RI is currently being realized in the European ERIGrid project that will support the technology development as well as the roll-out of smart grid solutions. It provides a holistic, CPES-based approach by integrating European research centres and institutions with outstanding lab infrastructure to jointly develop common methods, concepts, and procedures. In the following sections, the main idea behind everything and the corresponding research and development activities of the ERIGrid approach are explained.

\subsection{Overview and Approach}

\begin{figure}[!htbp]
	\centering
	\includegraphics[width=0.84\columnwidth]{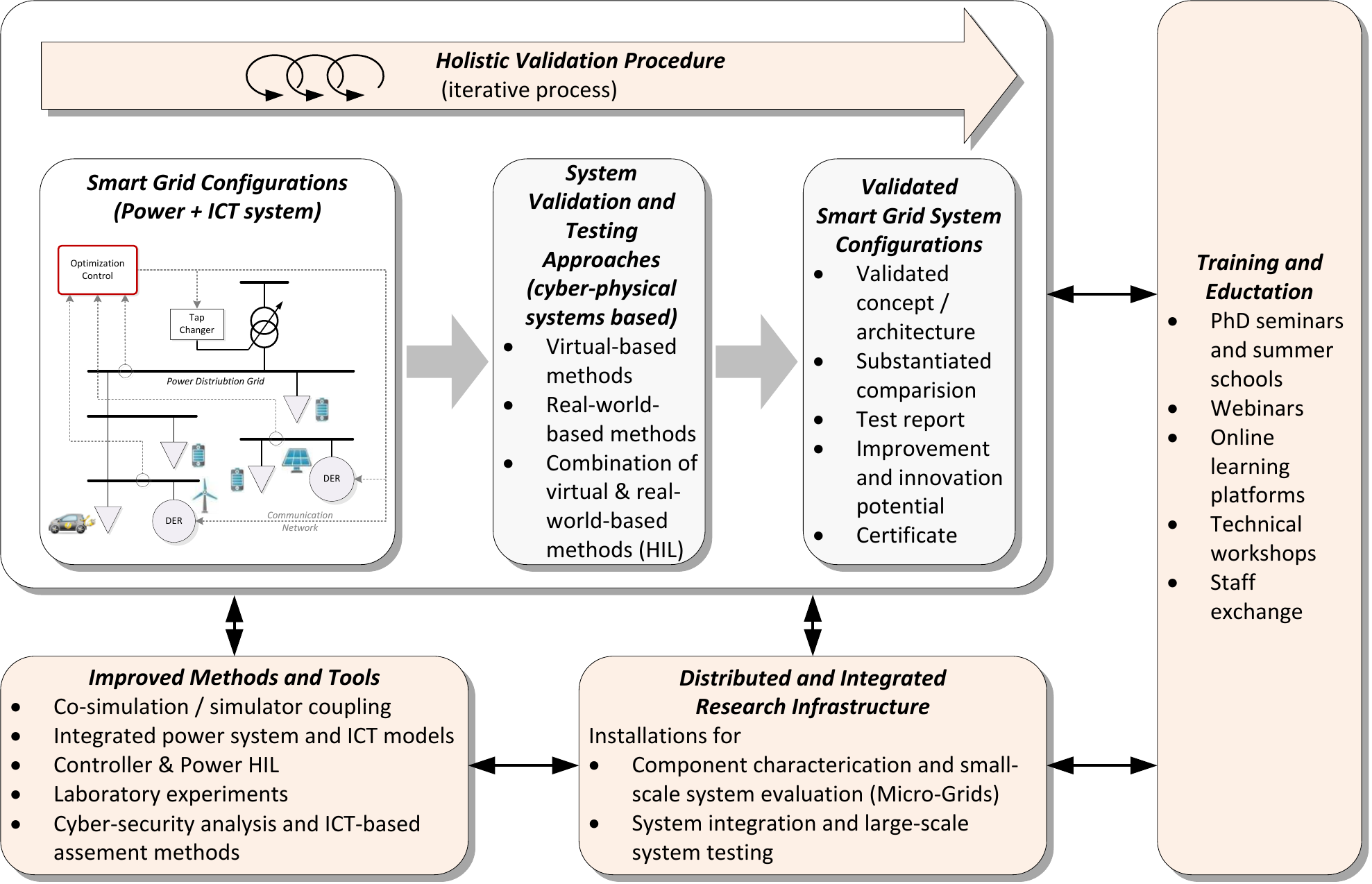}
	\caption{Overview of the ERIGrid approach}
	\label{fig:overiew}
\end{figure}

Figure~\ref{fig:overiew} provides an overview of the ERIGrid concept supporting the technology development, validation and roll out of smart grid solutions. The target of this integrating activity is to realise the systematic validation and testing of smart grid configurations from a holistic, cyber-physical systems point of view. It follows a multi-domain approach and covers power system, ICT and cyber-security topics in a cyber-physical manner. It is expected that the provision of support to the upcoming large-scale roll out of new concepts, technologies and approaches will also be possible fostering innovation.

The main research activities are related to the development of a formalized, holistic validation procedure, simulation and laboratory-based testing methods, and the provision of an integrated RI. Additionally, training and education concepts are also being developed. 

\subsection{Research and Development Directions}

\noindent \textit{a) Holistic Validation Procedure\\[-0.5em]}

\noindent Validating smart grid technologies and developments is a task that requires a holistic treatment of the overall process since the entire domain spectrum of CPES solutions has to be considered. This consideration has to be carried out alongside the technical components such as the grid infrastructure, storage, generation, consumption, etc. it also comprises customers, markets, ICT, regulation, governance, and metrology to name a few of them. It is clear that the full development process has to be covered. This includes design, analysis, testing, verification (even certification), as well as deployment. Furthermore, the whole range of aspects that are of interest and relevant to a stable, safe and efficient smart grid solutions have to be regarded. Thus, small-signal stability together with large-scale scenarios, short-term impacts and long-term sustainability, economic feasibility/profitability, and cyber-security have to be analysed. In fact, since all these topics are dependent upon each other, they have to be analysed in an integrated way. Finally, a holistic approach demands integrating all prospective R\&D sites and stakeholders, e.g. hardware/software simulation labs as well as academic and industrial research. \cite{Strasser:2017}    

\begin{figure}[!htbp]
	\centering
	\includegraphics[width=1.00\columnwidth]{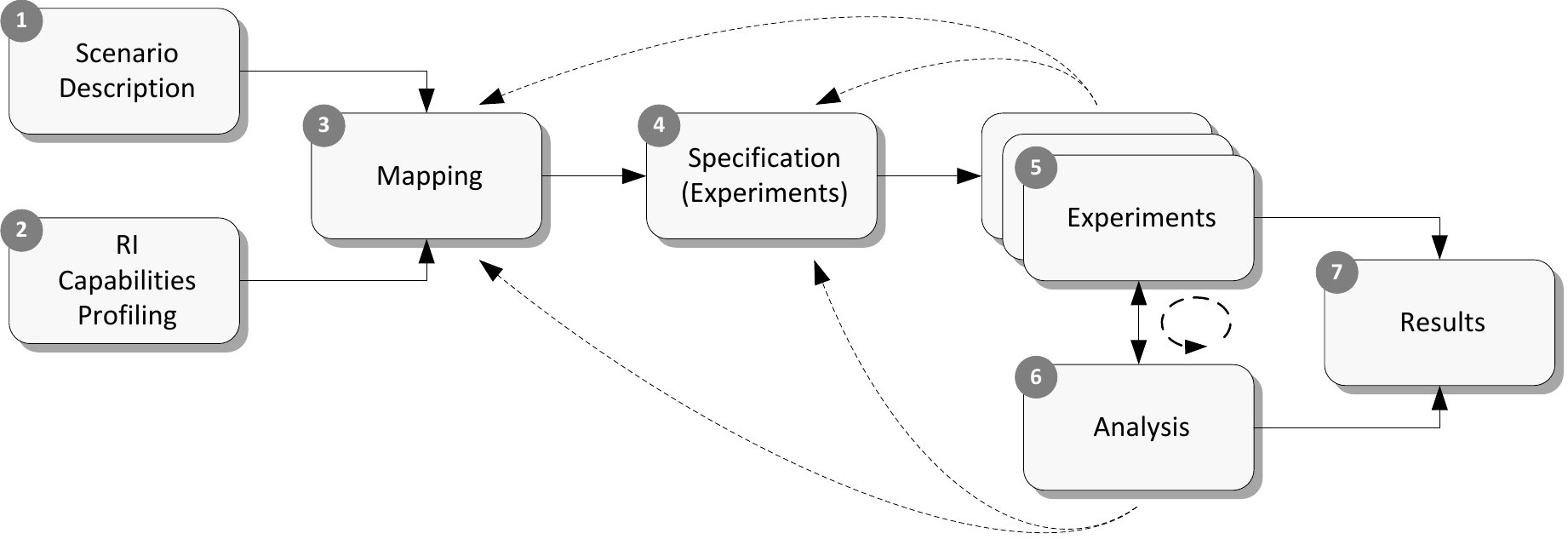}
	\caption{Different steps for carrying out a holistic validation of CPES}
	\label{fig:holistic_validation_procedure}
\end{figure}

Comparable processes have been successfully implemented in other application domains like automotive, consumer electronics, mechanical/chemical engineering (albeit on an arguably less complex level) \cite{Bringmann:2008,Fouchal:2016}. In order to realize a sustainable and cost effective holistic procedure in smart Grid system validation, two major challenges have to be addressed: \textit{(i)} formalized scenario design and \textit{(ii)} model exchange. These will ensure guaranteed comparability between experiments of different setups and designs. They will also facilitate subsequent re-utilization of experimental results from different stakeholders, as the basis for continuative experiments. Utilising the two aforementioned capabilities, use cases can then be defined with respect to the actual system requirements and with the best setup available. The huge amount of possible scenarios and experiments has to be narrowed down to a valid set of experiments in such a way that yields statistical aspects and reproducible results. Quantifying both the errors and significance of these experiments is of high importance. A consistent methodology for model exchange facilitates coordinated smart grid experiments (of representative scope and scale) within ERIGrid and beyond. It also aids the comparability between experiments, and furthermore to conduct consecutive, continuative, and parallel experiments for the re-utilization and subsequent use of results. In Figure~\ref{fig:holistic_validation_procedure}, the holistic validation process is outlined. \cite{Strasser:2017}  

In detail, the process is divided into the following seven steps \cite{Blank:2016}:

\begin{enumerate}
	\item \textit{Scenario Description}: It is important to decide on the system boundaries for each smart grid scenario. It imperative to define which parts of the overall system are being considered and to what level of detail. Thus, the use cases will be derived accordingly. The specific systems configuration will be described within each testcase .
	\item \textit{Research Infrastructure Capabilities Profiling}: In order to take advantage of the best suitable infrastructure setup for each use case, all ERIGrid installation (hardware and simulation-based labs) will be profiled. The capabilities of each lab installation will be analysed, documented and published. This will facilitates optimal use of capabilities provided by the existing hardware and software. 
	\item \textit{Mapping}: optimal mapping will be carried out based on both the scenario descriptions and partner profiles. The partner profiles will be analysed, with the goal of identifying which ERIGrid partner(s) is(are) best suitable for realizing each part of the defined use cases and especially the validation of the appropriate test cases. Particular attention will be paid to the interfaces between the various hardware and software components, while still ensuring that the mapping follows a cyber-physical systems approach.  
	\item \textit{Specification of Experiments:} After the determination of the best partner(s) setup (who), for the test cases (what), the concrete experiments will then be specified (how). In order to realize a use case the interfaces between the necessary hardware/software will be implemented based on the requirements imposed on the corresponding test cases. In addition, the available tool support will be analysed and taken into account. During this step missing models are expected to come to light, triggering the initialisation of their development. 
	\item \textit{Conducting Experiments:} After the experiments are specified they will be conducted following a consortium-wide specification. Here, it is highly important to identify the right amount and variations of experiments. This step is closely related to the next step addressing analysis tasks. 
	\item \textit{Analysis:} The step is interrelated to the previous step. Each experiment will be analysed and afterwards iteratively refined in order to meet the test case requirements. This cyclic process allows for the development of high quality experiments. Experimental results may be used to generate surrogate models of certain phenomena/aspects, that can be efficiently computated,  and may (if necessary in a virtual manner) be ``plugged into'' other partners’ setups.
	\item \textit{Test Results:} In the last step the final reports will be compiled as summaries and analysis of the experiments. Certificates will be given to components such as controllers that have been tested successfully against requirements that were defined in reference scenarios and systems. Based on experimental finding, improvement potentials as well as further innovation-related activities will be identified.
\end{enumerate}

\noindent\textit{b) Improved Methods and Tools\\[-0.5em]}

\noindent The current development in the smart grid field shows that future systems will contain a heterogeneous agglomeration of active power electronics and passive network components coupled via physical processes and dedicated communication connections to automation systems (SCADA, DMS) \cite{Strasser:2017}. In order to analyse and evaluate such a multi-domain configurations, a set of corresponding methods and tools is necessary. In the context of smart grids the following possibilities exist \cite{Strasser:2017}:

\begin{itemize}
	\item Pure virtual-based methods and tools (mainly simulation-based),
	\item Real-world-based methods (laboratory and demonstration-based), or
	\item A combination of both (hardware-in-the-loop based).
\end{itemize}

\begin{figure}[!htbp]
	\centering
	\includegraphics[width=0.89\columnwidth]{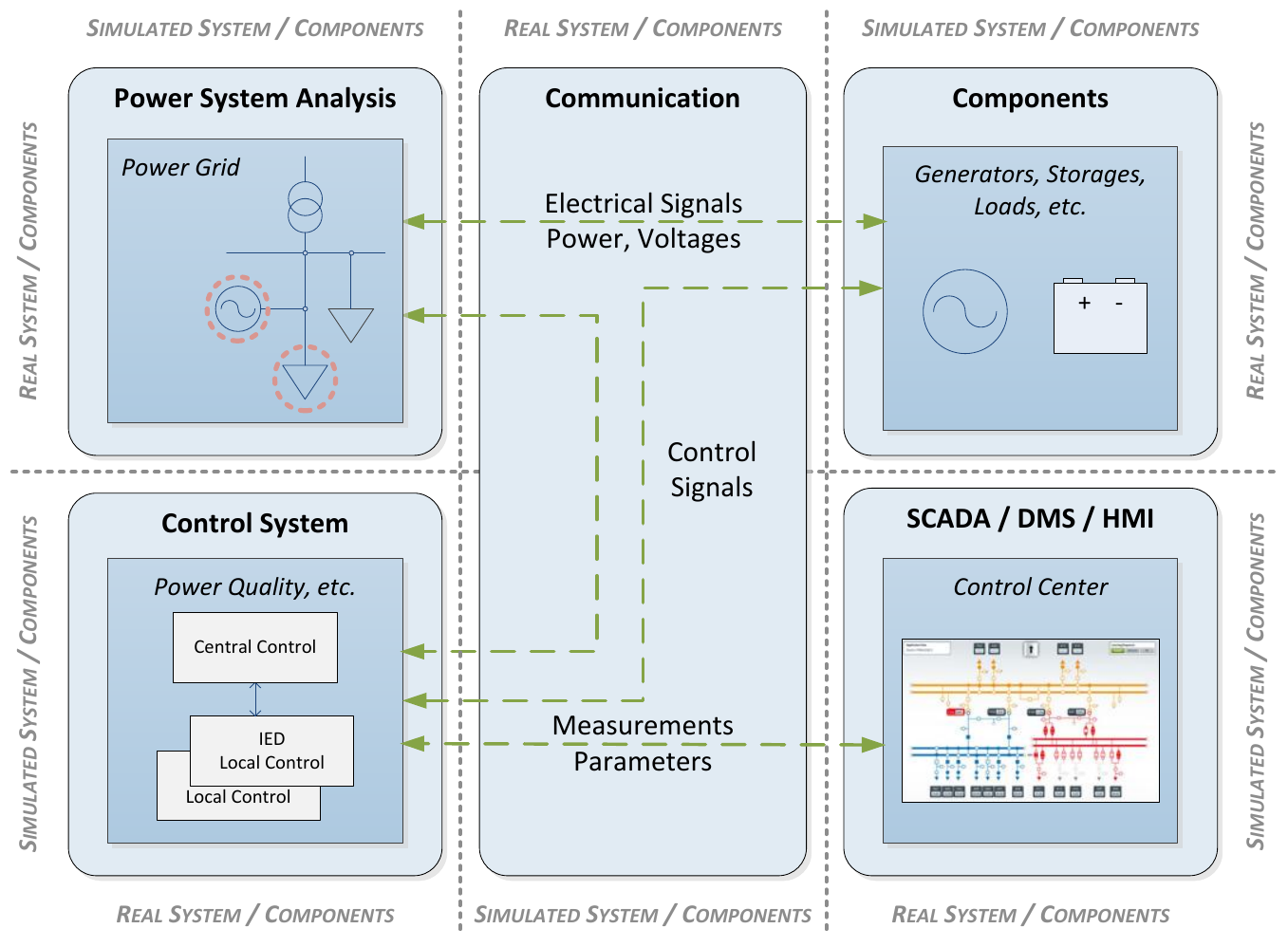}
	\caption{Improved methods and tools for smart grid validation -- possibility to combine virtual (simulated) and real components \cite{Strasser:2017}
	}
	\label{fig:improved_methods}
\end{figure}

Figure~\ref{fig:improved_methods} provides an overview of the methods and tools which will be applied in ERIGrid. The goal is to use all three above mentioned possibilities in order to cover the whole range of opportunities in a cyber-physical manner that is essential for addressing system integration and validation questions. Usually, pure virtual-based methods are not always sufficient for validating CPES since the availability of proper and accurate simulation models cannot always be guaranteed (e.g., inverter-based components are sometimes very complex to model or it takes too long to get a proper model). Also laboratory-based testing approaches and a combination of both (hardware-in-the-loop based) have to be applied. 

ERIGrid follows an approach where the following methods, models and corresponding tools are being improved for component characterization, system integration and validation:

\begin{itemize}
	\item Co-simulation (simulator couplings covering the power system, components, communication, control system and SCADA),
	\item Integrated power system and ICT models,
	\item Controller \& Power Hardware-in-the-Loop (HIL),
	\item Laboratory experiments, and
	\item Cyber-security analysis and ICT-based assessment methods.
\end{itemize}

It is usually not possible to test the whole power system infrastructure in a lab environment and also pure simulation-based methods are sometimes not precise enough. A very important issue in ERIGrid approach is that parts of the smart grid system can be available as a physical component (in a laboratory environment), as a simulation model or even as a combination of both. 

Further research is necessary in order to combine all three validation approaches in a flexible manner (based on the testing needs of a specific CPES application). In the ERIGrid context this includes the coupling of software tools and simulators, the model exchange between them as well as the provision of necessary component models. The coupling of simulation systems with laboratory equipment is also in focus so as to facilitate the above outlined flexible interaction between hardware and software components (see Figure~\ref{fig:improved_methods}). \\

\noindent\textit{c) Distributed and Integrated Research Infrastructure\\[-0.5em]}

\noindent Historically, power systems have been nationally organized, however, the main vision of the ERIGrid approach is to bring those institutions together on European level. 

Key efforts will be applied in the realization of a distributed and integrated RI which is capable to support the validation and testing of diverse CPES configurations. The power and energy systems domain is characterized by diverse lab environments operated by universities, research centres, and industry in various European countries. As these RIs deal with many levels of system operation, their components and ICT systems are very heterogeneous. There is no central European RI for smart grid research, and harmonization efforts have till now focussed on networking activities. The ERIGrid vision extends beyond networking to allow further integration and effective use of testing resources through human interoperation and mutual access to a shared experimental platform.

Stronger interoperability is achieved by means of three main efforts: \textit{(a)} a harmonized and detailed description of the available RI aimed at supporting the collaborative design of experiment configurations, \textit{(b)} the development of a harmonized validation procedure that accommodates the integration of testing procedures and resources across RIs, facilitating the test design and evaluation process, and \textit{(c)} a coherent and technology-agnostic description of RI control interfaces, supporting the coordination of control software deployment across RIs. The harmonized RI description \textit{(a)} will soon be available as a publicly accessible database whereas the harmonized validation procedure \textit{(b)} is still under development, but has been outlined in \cite{Blank:2016}.

\begin{figure}[!htbp]
	\centering
	\includegraphics[width=0.83\columnwidth]{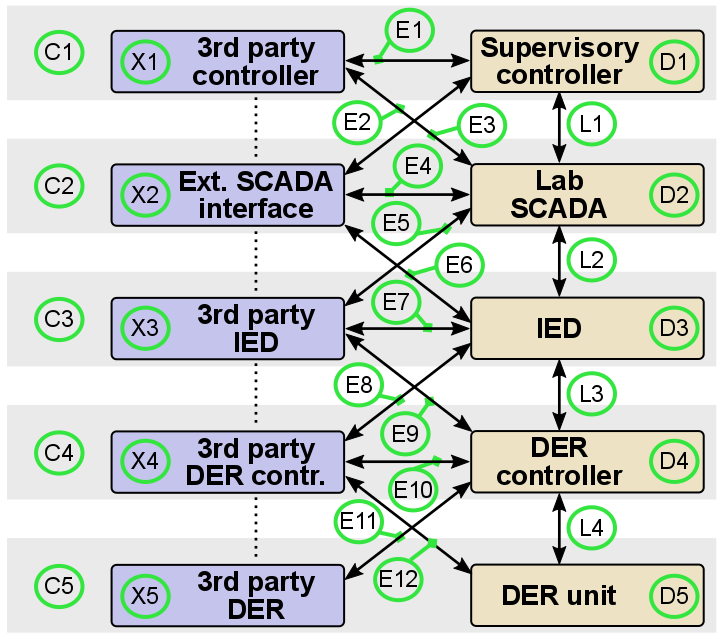}
	\caption{Harmonized control interfaces}
	\label{fig:harmonized_control_interfaces}
\end{figure}

The coherent RI control interface description \textit{(c)} resulted in a taxonomy as illustrated in Figure~\ref{fig:harmonized_control_interfaces}. The five control levels are marked C1~\dots~C5. Local controllers have been assigned the levels D1~\dots~D5, corresponding to their associated control levels, while external controllers are denoted X1~\dots~X5. Communication between the local controllers, i.e. communication links which are considered to be part of the laboratory infrastructure, are enumerated L1~\dots~L4, while those communication links which enable the interaction between the laboratory and external controllers, have been assigned the identifiers E1~\dots~E12. This taxonomy establishes a combination of structural and functional criteria that can be applied across smart grid laboratories to map the local architecture into a common framework, much like the Smart Grid Architecture Model (SGAM) plane serves that purpose for smart grid standardisation \cite{Uslar:2012}. 

This model outlines a number of possible interactions for external control software deployment, six of which are of particular interest for CPES scenarios:

\begin{enumerate}
\item Communication interfaces provided by a lab-internal DER controller, allowing an external Intelligent Electronic Device (IED) to control the DER (E11). 
\item Communication interfaces provided by a lab-internal IED, allowing an external IED to influence the behaviour of the internal IED (E7).
\item Communication interfaces provided by a lab-internal IED, allowing the IED to be remotely operated through an external SCADA system (E6).
\item Communication interfaces provided by a lab-internal SCADA system, allowing an external SCADA system to receive and/or send data to laboratory devices through the internal SCADA system (E4).
\item Communication interfaces provided by a lab-internal SCADA system to an external controller (E3).
\item Communication interfaces provided by a lab-internal supervisory controller to an external controller (E1), for example to allow the external controller to influence the control behaviour of the supervisory controller.
\end{enumerate}

Other interactions are possible, however they are considered to be of less interest for the ERIGrid case. \\

\noindent\textit{d) Education and Training\\[-0.5em]}

\noindent Drivers for the transition to a CPES are largely due to the rise in renewable energy sources, e-mobility, distributed energy storages, more resilient network infrastructure, new market models and the goal of an emission free and sustainable energy system \cite{EC:2007,EC:2011,IEA:2011}. The need to understand the interconnections between the power system components increases steadily due to the rising complexity caused by the myriad of players and actors. This includes not only the physical system but also the communication and the control of connected power components and assets. These needs are addressed by recent developments and research in cyber-physical systems \cite{Strasser:2015}. As a result of this trend a further focus in ERIGrid needs to be put on educational aspects covering the transition in power systems towards a smart grid. 

Training and education can be based on the concepts of modelling and simulating components as well as on conducting laboratory experiments to understand how they work as a system. When designing intelligent automation and control concepts like energy management systems, voltage control algorithms, dynamic protection, topology re-configuration as well as demand response mechanisms, researchers and engineers need to understand different control paradigms, like centralized, hierarchical and distributed approaches. Moreover, they also have to learn  how to use and be aware of different tools; their strengths and weaknesses, and how to interconnect them. 

The goal is to design proper training material as well as the necessary procedures for educating power system and ICT professionals, (young) researchers as well as students in the domain of smart grid system integration and validation.
%
%
\section{Conclusions and Outlook}
\label{sec:conclusions}

A large-scale roll out of smart grid solutions and corresponding technologies and products is expected in the upcoming years. The validation of such complex CPES needs attention since available testing approaches and methods are mainly focusing on power system and ICT components. An integrated, cyber-physical systems based, multi-domain approach for a holistic testing of smart grid solutions is currently still missing.

Four main research priorities have been identified and prioritized in context of the European project ERIGrid in order to overcome the shortcomings in the validation of today's smart grid systems. The focus of the future research lies in the development of a holistic validation procedures and corresponding formalized test case descriptions as well as the improvement of simulation-based methods, hardware-in-the-loop approaches, and lab-based testing (incl. a flexible combination of them). In addition, the integration and harmonization of nationally organized labs into a distributed and integrated RI is a further priority. Last but not least, the education and training of researchers and power system professionals on these extended and harmonized testing possibilities is another field of action. 
%
%
\subsubsection*{Acknowledgments.} This work is supported by the European Community’s Horizon 2020 Program (H2020/2014-2020) under project ``ERIGrid'' (Grant Agreement No. 654113). 
%
%
\bibliographystyle{splncs03}
\bibliography{literature}
\end{document}